\documentclass[twocolumn,superscriptaddress,showpacs,preprintnumbers,amsmath,amssymb,prd]{revtex4}
\usepackage{graphicx}
\usepackage{amsmath}
\usepackage{color}
\UseRawInputEncoding

\begin{document}

\thispagestyle{empty}

\title{On the convergence of the polarization tensor in space-time of three dimensions}

\author{M.~Bordag}
\affiliation{Institut f\"{u}r Theoretische Physik, Universit\"{a}t Leipzig,
D-04081, Leipzig, Germany
}
\affiliation{Bogoliubov Laboratory of Theoretical Physics,
Joint Institute for Nuclear Research, 141980, Dubna, Russia}

\author{
G.~L.~Klimchitskaya}
\affiliation{Central Astronomical Observatory at Pulkovo of the Russian Academy of Sciences, St.Petersburg,
196140, Russia}
\affiliation{Peter the Great Saint Petersburg
Polytechnic University, Saint Petersburg, 195251, Russia}

\author{
V.~M.~Mostepanenko}
\affiliation{Central Astronomical Observatory at Pulkovo of the Russian Academy of Sciences, St.Petersburg,
196140, Russia}
\affiliation{Peter the Great Saint Petersburg
Polytechnic University, Saint Petersburg, 195251, Russia}
\affiliation{Kazan Federal University, Kazan, 420008, Russia}

\begin{abstract}
In this paper, we consider the convergence properties of the polarization
tensor of graphene obtained in the framework of thermal quantum field theory
in three-dimensional space-time. During the last years, this problem attracted
much attention in connection with calculation of the Casimir force in graphene
systems and investigation of the electrical conductivity and reflectance of
graphene sheets. There are contradictory statements in the literature, especially
on  whether this tensor has an ultraviolet divergence in three dimensions.
Here, we analyze this problem using the well known method of dimensional
regularization. It is shown that the
thermal correction to the polarization tensor is finite at any $D$, whereas its
zero-temperature part behaves differently for $D=3$ and 4. For $D=3$, it is
obtained by the analytic continuation with no
subtracting infinitely large terms. As to the space-time of $D=4$,
the finite result for the polarization tensor at zero temperature is found after
subtracting the pole term. Our results are in agreement with previous calculations
of the polarization tensor at both zero and nonzero temperature. This opens
possibility for a wider application of the quantum field theoretical approach in
investigations of graphene and other two-dimensional novel materials.
\end{abstract}

\maketitle

\section{Introduction}
\newcommand{\nn}{\nonumber}\renewcommand{\ni}{\noindent}
\newcommand{\eq}[2]{\begin{align}\label{#1}#2\end{align}}
\newcommand{\eqo}[1]{\begin{align*}#1\end{align*}}
\newcommand{\Tr}{{\rm Tr}}
\newcommand{\pa}{\partial}
\renewcommand{\t}{\mathbf{t}}
\newcommand{\q}{\mbox{\boldmath${q}$}}
\newcommand{\vk}{\mbox{\boldmath${k}$}}
\newcommand{\A}{{\cal A}}\newcommand{\F}{{\cal F}}\renewcommand{\L}{{\cal L}}
\newcommand{\ep}{\epsilon}
\newcommand{\ve}{\varepsilon}
\newcommand{\la}{\lambda}
\newcommand{\om}{\omega}\newcommand{\al}{\alpha}
\newcommand{\be}{\beta}\newcommand{\ga}{\gamma}
\newcommand{\vphi}{\varphi}
\renewcommand{\v}{v_{\rm F}}
\newcommand{\tk}{\mbox{$\tilde{k}$}}
\newcommand{\tvk}{\mbox{\boldmath$\tilde{k}$}}
\newcommand{\zmn}{Z^{\mu\nu}}
\newcommand{\tzmn}{\widetilde{Z}^{\mu^{\prime}\nu^{\prime}}}
\newcommand{\ozmn}{\widetilde{Z}_1^{\mu^{\prime}\nu^{\prime}}}
\newcommand{\gmn}{g^{\mu^{\prime}\nu^{\prime}}}

The term polarization tensor has many different meanings and was used for theoretical
description of diverse physical phenomena. Here, we reconsider the problem of
convergence  of the vacuum photon polarization tensor of graphene in quantum electrodynamics (QED)
at nonzero temperature in three-dimensional space-time. Independently of an entirely theoretical
interest to calculation of the polarization tensor at both zero and nonzero
temperature for the case of $D=(2+1)$ dimensions \cite{0,1,2,3,4}, this
problem attracted special attention \cite{5,6,7,8,9}
in connection with the advent of two-dimensional hexagonal structure of
carbon atoms called graphene \cite{10}.

At energies below a few eV, the electronic properties of graphene are well-described by
a set of massless or very
light quasiparticles with spin 1/2 obeying the Dirac equation, where the speed
of light $c$ is replaced with the Fermi velocity $\v\approx c/300$ \cite{11,12,13,14}
(in the following text, we use the system of units where $\hbar=c=1$). This has opened an
attractive opportunity of describing the reaction of graphene to the electromagnetic
field using the well established methods of QED in (2+1) dimensions, especially
the concept of the polarization tensor, i.e., restricting to the one-loop radiative
correction in the language of QED.
Taking into account that the properties of
graphene strongly depend on temperature, this well may be done in the framework of
thermal quantum field theory.

The polarization tensor derived in (2+1)-dimensional quantum field theory (QFT) \cite{0,1}
was first applied for theoretical description of the Casimir force between two
graphene sheets in Ref.~\cite{15}. In Ref.~\cite{16} this tensor was generalized
for the case of nonzero temperature and calculated at the pure imaginary Matsubara
frequencies taking into account the nonzero mass of quasiparticles and chemical potential.
The obtained results were used for investigation of the Casimir and Casimir-Polder
forces in various configurations \cite{17,18,19,20,21,22,25,26,27,28,29,29a}.

The analytic continuation of the polarization tensor of graphene to the entire
plane of complex frequencies, including the real frequency axis, was performed in
Ref.~\cite{30}. These results were generalized for the graphene
sheets possessing a nonzero chemical potential \cite{31}. The obtained polarization
tensor of graphene at nonzero temperature was used in calculation of the Casimir
and Casimir-Polder forces in graphene systems \cite{32,34,35,36,37,38,39,40,41,42},
electrical conductivity \cite{43,44,45,46} and reflectivity properties of
graphene \cite{30,47,48,49}. Computations of the Casimir force in graphene
systems using the polarization tensor have been found to be in excellent
agreement with the measurement data of two precision experiments \cite{51,52,53,54}.

In spite of big progress in application of thermal QFT for obtaining
the polarization tensor of graphene and describing its properties on this basis, the
more phenomenological theoretical approach using the Kubo formula is often used
in the literature for the same purpose (see, e.g.,
Refs.~\cite{55,56,57,58,59,60,61,62,63,64,65,66,67,68,69,70,71,72,73,74}).
There are, however, significant conceptual differences between the quantum field
theoretical and Kubo approaches. For instance, in the framework of the Kubo approach,
dissipation is introduced by means of the phenomenological relaxation parameter
treated as the imaginary part of complex frequency. Alternatively, the QFT
does not use phenomenological parameters and describes dissipation by
means of the imaginary part of the polarization tensor which arises for the
scaled 3-momentum magnitudes exceeding the energy gap in graphene.

In the spatially local approximation, there is an agreement between the results
obtained using different theoretical approaches \cite{20,30,43,44,45,47,48,49}.
As to the spatially nonlocal case, the quantum field theoretical approach predicts
the presence of a double pole at zero frequency in the transverse dielectric
permittivity of graphene \cite{42}, which is not obtainable in the Kubo approach.
It was stated \cite{75} that the presence of a double pole might be connected
with an improper regularization of the polarization tensor obtained within thermal
QFT.

In this regard, it should be noted that there are contradictory statements in recent
literature concerning the convergence of this tensor. Thus, Ref.~\cite{16} found
by power counting
that the polarization tensor of graphene diverges and made it finite by a
{Pauli-Villars}
subtraction, whereas Refs.~\cite{28,29,30} conclude
that in 2+1 dimensions it is finite because the ultraviolet divergence is not present
due to the gauge invariance. As to Ref.~\cite{75}, it states that the polarization
tensor of
graphene obtained by means of the quantum field theory is divergent and suggests
an alternative regularization procedure, which brings it to exact coincidence with
that obtained by means of the Kubo approach.

In view of the above, we feel that it is necessary to clarify the situation.
We demonstrate the calculation of the polarization tensor of graphene using the
methods of QED at nonzero temperature in detail and in such a way, that the
calculation can be followed with a minimum of knowledge of the field theoretical
methods. Thereby it must be underlined that in the standard QED at zero temperature
the polarization tensor was calculated long ago both in (3+1) dimensions (see, e.g.,
the textbooks \cite{75a,80}) and in (2+1) dimensions \cite{0,1}. For the latter case,
the key moments, gauge invariance and ultraviolet finiteness were mentioned
explicitly in Sec.~2 of Ref.~\cite{1}.

In the present paper, we reconsider the polarization tensor appearing in the quantum
field theoretical approach to graphene at nonzero temperature in detail. We use
dimensional regularization. First we demonstrate how the transversality of the
polarization tensor can be seen before the momentum integration.
Next, we demonstrate that this tensor consists of the zero-temperature part and a
thermal correction to it. The immediate analytic calculation shows that the thermal
correction to the polarization tensor is finite, so that the ultraviolet
divergence, if any, might be contained only in its zero-temperature part.
Then, we use the exponential representation for the propagators and carry out the
momentum integrations. Finally, after carrying out the next-to-last integration,
the transversality becomes evident also in this representation as well as the ultraviolet
properties.

Specifically, by considering the polarization
tensor in the space-time of complex $D$ dimensions, we demonstrate that in
the case of $D=3$ the finite result is obtained
{using the regularization by means of}
analytic continuation from the case ${\rm Re}D<2$.
In so doing, no pole terms need to be subtracted,
{ i.e., no renormalization is needed}.
Applying the same procedure to
the polarization tensor in the space-time of $D=4$ dimensions, we show that for
obtaining the finite result it is necessary to subtract the pole term,
{ i.e., regularization should be followed by the renormalization.}
Generally speaking,
such behavior in the ultraviolet region is well known in QFT as a consequence of,
together, power counting, gauge invariance, and parity. However, an active discussion
for graphene, which exists in two spatial dimensions but interacts with the
electromagnetic field existing in three-dimensional space, revealed the necessity to
demonstrate this behavior in detail. The performed analysis is
in confirmation of the polarization tensor derived in the literature in the
framework of both ordinary and thermal QFT.

The paper is organized as follows. In Section~II, we consider general expression
for the polarization tensor of graphene at nonzero temperature in the space-time
of $D$ dimensions. Section~III is devoted to the zero-temperature part of the
polarization tensor and its analytic properties.
In Section~IV,  the convergence properties of the polarization tensor in both
three- and four-dimensional space-time are considered. In Section~V, the reader
will find our conclusions and a discussion.

Recall that use the system of units where $\hbar=c=1$.

\section{Representation of the polarization tensor at nonzero
temperature in $D$ dimensions}

\begin{figure}[!t]
\vspace*{-5cm}
\hspace*{-1cm}
\includegraphics[width=3.0in]{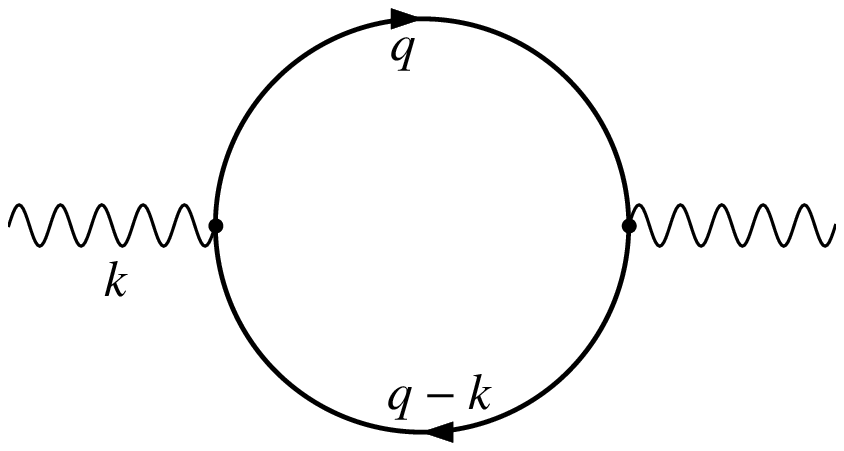}
\vspace*{-3cm}
\caption{\label{fg1}
Feynman diagram representing the one-loop polarization tensor of
graphene. }
\end{figure}

In the framework of QFT the one-loop polarization tensor of
graphene was considered in many papers (see, e.g., Refs.~\cite{8,15,16,17,30,31,36,76}).
It is represented by a simple diagram of Fig.~\ref{fg1} where the solid lines depict
the propagators of fermionic quasiparticles which move with the Fermi velocity $\v$
and satisfy the Dirac equation in 2+1 dimensions
\begin{eqnarray}
&&
\left[\ga^0\left(i\frac{\pa}{\pa t}-eA_0\right)+\tilde{\ga}^1\left(
i\frac{\pa}{\pa x^1}-eA_1\right)\right.
\nonumber \\
&&~~~~
\left.+\tilde{\ga}^2\left(
i\frac{\pa}{\pa x^2}-eA_2\right)-m{\v}^2\right]\psi(x)=0.
\label{eq1}
\end{eqnarray}
\noindent
Here, $\ga^{\nu}$ are the standard Dirac matrices, $\tilde{\ga}^{1,2}=\v\ga^{1,2}$,
$A_{\nu}=(A_0,A_1,A_2)$ is the vector potential of the electromagnetic field, and
$m$ is the mass of quasiparticles bearing the electric charge $e$.

The important feature of Eq.~(\ref{eq1}) is that the interaction of charged
quasiparticles with the electromagnetic field is introduced by the standard substitution
\begin{equation}
i\frac{\pa}{\pa x^{\nu}}\longrightarrow i\frac{\pa}{\pa x^{\nu}}-eA_{\nu},
\label{eq2}
\end{equation}
\noindent
where, for a graphene sheet in the plane $x^3=0$, it holds $x^{\nu}=(t,x^1,x^2,0)$,
$\nu=0,\,1,\,2,\,3$. Note that Eq.~(\ref{eq2})
contains the speed of light in the factor $e/c$ (we recall that here $c=1$).
This reflects the fact that the electromagnetic field, although it interacts with the
quasiparticles confined in a graphene plane, exists in the 3+1 dimensional bulk.
As a consequence, in the Dirac model of graphene, the electric charge in the system
of units with $\hbar=c=1$ is not dimensional (as it holds in the strictly 2+1
dimensional electrodynamics \cite{0}) but dimensionless and results in the standard
fine structure constant $\al=e^2\approx1/137$.

Calculation of the diagram shown in Fig.~\ref{fg1} includes an integration over
the internal momentum $q=(q_0,\q)$ and taking the trace of $\gamma$-matrices (see
Refs.~\cite{16,30} for details). Keeping in mind that we are looking for the
polarization tensor of graphene at any temperature $T$, within the Matsubara formalism,
an integration over $q_0$ should be replaced with a summation over the pure imaginary
fermionic Matsubara frequencies
\begin{equation}
q_{0n}\equiv iq_{Dn}=2\pi ik_BT\left(n+\frac{1}{2}\right),
\label{eq3}
\end{equation}
\noindent
where $n=0,\,\pm1,\,\pm2,\,\ldots$, and $k_B$ is the Boltzmann constant.
In so doing, the zero component of the external, photon, wave vector $k=(k_0,\vk)$
is equal to the pure imaginary bosonic Matsubara frequencies
\begin{equation}
k_{0n}\equiv ik_{Dl}=2\pi ik_BTl.
\label{eq4}
\end{equation}

Although here and below we deal with graphene, which is the two-dimensional sheet of
carbon atoms, in the following we use the $D$-dimensional vectors
$(q_0,\q)=(q_0,\,q^1,\,\ldots,\,q^{D-1})$ and $(k_0,\vk)=(k_0,k^1,\,\ldots,\,k^{D-1})$,
where the dimension of the spatial part is $D-1$, and respective integration measures.
The metric tensor is defined as $g_{\mu\nu}={\rm diag}(1,-1,-1,\,\ldots,\,-1)$
and the product of two vectors is $qk=q_{\nu}k^{\nu}=q_0k^0-\q\vk$.
The trace of the metric tensor is $g_{\nu}^{\,\nu}=D$.
The point is that, in general, the polarization tensor is ultraviolet divergent
like most radiative corrections in QFT. For instance, simple power counting shows a
divergence also in (2+1)-dimensions. For this reason, a regularization is
necessary. By introducing the $D$-dimensional space-time, we take the dimensional
regularization which amounts in formally taking a complex dimension $D$ (see,
e.g., Sec.~11.2 in the textbook \cite{76a}).
This allows to find the analytic properties of the polarization
tensor as the function of $D$.

Note  that for graphene the Dirac cones are located at the two
points at the corners of the Brillouin zone \cite{12}.
Then, after taking the trace over the gamma matrices,
the resulting
polarization tensor in the momentum representation is given by \cite{30}
\begin{eqnarray}
&&
\Pi^{\mu\nu}(ik_{Dl},\vk,T)=-\frac{32\pi\al}{\v^2}\,k_BT
\label{eq5} \\
&&~~~~~~\times
\sum_{n=-\infty}^{\infty}
\int\frac{d^{D-1}\q}{(2\pi)^{D-1}}\,
\frac{\zmn(ik_{Dl},\vk;iq_{Dn},\q)}{R(ik_{Dl},\vk;iq_{Dn},\q)},
\nonumber
\end{eqnarray}
\noindent
where
\begin{equation}
\zmn(ik_{Dl},\vk;iq_{Dn},\q)=\eta_{\mu^{\prime}}^{\,\mu}\eta_{\nu^{\prime}}^{\,\nu}
\tzmn(ik_{Dl},\vk;iq_{Dn},\q)
\label{eq6}
\end{equation}
\noindent
and $\eta_{\mu}^{\,\nu}={\rm diag}(1,\v,\,\v,\,\ldots,\,\v)$.

The quantities $\tzmn$ and $R$ are
\begin{eqnarray}
&&
\tzmn(ik_{Dl},\vk;iq_{Dn},\q)=q^{\mu^{\prime}}(q-\tk)^{\nu^{\prime}}+
(q-\tk)^{\mu^{\prime}}q^{\nu^{\prime}}
\nonumber \\
&&~~~~~~~~
+\gmn[-q(q-\tk)+m^2],
\label{eq7} \\
&&
R(ik_{Dl},\vk;iq_{Dn},\q)=(q^2-m^2+i0)
\nonumber \\
&&~~~~~~~~~
\times[(q-\tk)^2-m^2+i0],
\nonumber
\end{eqnarray}
\noindent
where  the infinitely small additions $i0$ originate from the fermion propagators,
the scaled momentum is
$\tk=(k_0,\v\vk)$, $q_0=q_{0n}=iq_{Dn}$ and $\tk_0=k_{0l}=ik_{Dl}$ in
accordance with Eqs.~(\ref{eq3}) and (\ref{eq4}). For instance,
\begin{eqnarray}
&&
\tilde{Z}^{00}(ik_{Dl},\vk;iq_{Dn},\q)=-q_{Dn}(q_{Dn}-k_{Dl})
\nonumber \\
&&~~~~~~~~~~~~~
+\q(\q-\tvk)+m^2,
\nonumber \\
&&
\tilde{Z}^{11}(ik_{Dl},\vk;iq_{Dn},\q)=2q^1(q^1-\tk^1)-q_{Dn}(q_{Dn}-k_{Dl})
\nonumber \\
&&~~~~~~~~~~~~~
-\q(\q-\tvk)-m^2
\label{eq8}
\end{eqnarray}
\noindent
etc., and
\begin{eqnarray}
&&
R(ik_{Dl},\vk;iq_{Dn},\q)=\left[q_{Dn}^2+\Gamma^2(\q)-i0\right]
\nonumber \\
&&~~~~~~~~~
\times\left[(q_{Dn}-k_{Dl})^2+\tilde{\Gamma}^2(\q,\vk)-i0\right],
\label{eq9}
\end{eqnarray}
\noindent
where
\begin{equation}
\Gamma^2(\q)=\q^2+m^2, \qquad
\tilde{\Gamma}^2(\q,\vk)=(\q-\tvk)^2+m^2.
\label{eq10}
\end{equation}

It is  common knowledge that electrodynamics is the gauge invariant theory.
This means that the Fourier transformed vacuum current
\begin{equation}
J^{\nu}(k)=\Pi^{\mu\nu}A_{\mu}(k)
\label{eq11}
\end{equation}
\noindent
should be invariant under the gauge transformation
\begin{equation}
\delta A_{\mu}(k)=\tilde{A}_{\mu}(k)-A_{\mu}(k)=ik_{\mu}\chi(k),
\label{eq12}
\end{equation}
\noindent
where $\chi(k)$ is an arbitrary function \cite{77}. As a consequence,
\begin{equation}
\delta J^{\nu}(k)=ik_{\mu}\Pi^{\mu\nu}\chi(k)=0.
\label{eq13}
\end{equation}

Thus, for the polarization tensor, the gauge invariance is realized in the form
of the transversality condition
\begin{equation}
k_{\mu}\Pi^{\mu\nu}=0.
\label{eq14}
\end{equation}

It is easily seen that the polarization tensor of graphene (\ref{eq5}) satisfies
this condition like that in full QED.
Really, using Eqs.~(\ref{eq6}) and (\ref{eq7}), by a simple rewriting, one obtains
\begin{eqnarray}
&&
k_{\mu}Z^{\mu\nu}=\v[2q^{\nu}\tk q-q^{\nu}\tk^2-\tk^{\nu}q^2+m^2\tk^{\nu}]
\label{eq15} \\
&&~~~~~~
=\v\left\{(q^2-m^2)(q-\tk)^{\nu}-[(q-\tk)^2-m^2]q^{\nu}\right\},
\nonumber
\end{eqnarray}
\noindent
{where $\tk^{\nu}=\eta_{\beta}^{\,\,\nu}k^{\beta}$.}

Then, from Eqs.~(\ref{eq5}) and (\ref{eq15}) we find
\begin{eqnarray}
&&
k_{\mu}\Pi^{\mu\nu}=-\frac{32\pi\al}{\v}\,k_BT\sum_{n=-\infty}^{\infty}\int
\frac{d^{D-1}\q}{(2\pi)^{D-1}}
\nonumber \\
&&~~~
\times\left[
\frac{(q-\tk)^{\nu}}{(q-\tk)^2-m^2}
-\frac{q^{\nu}}{q^2-m^2}\right].
\label{eq16}
\end{eqnarray}
\noindent
{Note that $q^0=q_{0n}=iq_{Dn}$ given by Eq.~(\ref{eq3}).

The integral in Eq.(\ref{eq16}) converges under the condition ${\rm Re}D<2$.
Using this condition, the seemingly divergent integral/sum is regularized allowing
the shift of variables $q\to Q+\tk$ where
\begin{eqnarray}
&&
\mbox{\boldmath{$Q$}}=\mbox{\boldmath{$q$}}-\mbox{\boldmath{\tk}},
\nonumber\\
&&
Q_0=q_0-\tk_0=q_0-k_0=i(q_{Dn}-k_{Dl})
\label{eq17}
\end{eqnarray}
\noindent
and $q_{Dn},{\ }k_{Dl}$ are defined in Eqs.~(\ref{eq3}) and (\ref{eq4}).
As a result,  the integrand itself vanishes,}
 i.e., the polarization tensor (\ref{eq5}) satisfies the transversality
condition (\ref{eq14}) even before carrying out the momentum integration.

Now we represent the polarization tensor (\ref{eq5}) as the part, which is independent
on temperature, and the thermal correction to it. For this purpose, the right-hand of
Eq.~(\ref{eq5}) is rewritten as
\begin{equation}
k_BT\sum_{n=-\infty}^{\infty}
f(ik_{Dl},\vk;iq_{Dn}),
\label{eq18}
\end{equation}
\noindent
where
\begin{equation}
f(ik_{Dl},\vk;iq_{Dn})=-\frac{32\pi\al}{\v^2}\int\frac{d^{D-1}\q}{(2\pi)^{D-1}}
\,\frac{Z^{\mu\nu}(ik_{Dl},\vk;iq_{Dn},\q)}{R(ik_{Dl},\vk;iq_{Dn},\q)}
\label{eq19}
\end{equation}
\noindent
(below we omit the already specified repeated arguments).

\begin{figure}[!t]
\vspace*{-4.cm}
\hspace*{-2cm}
\includegraphics[width=4.0in]{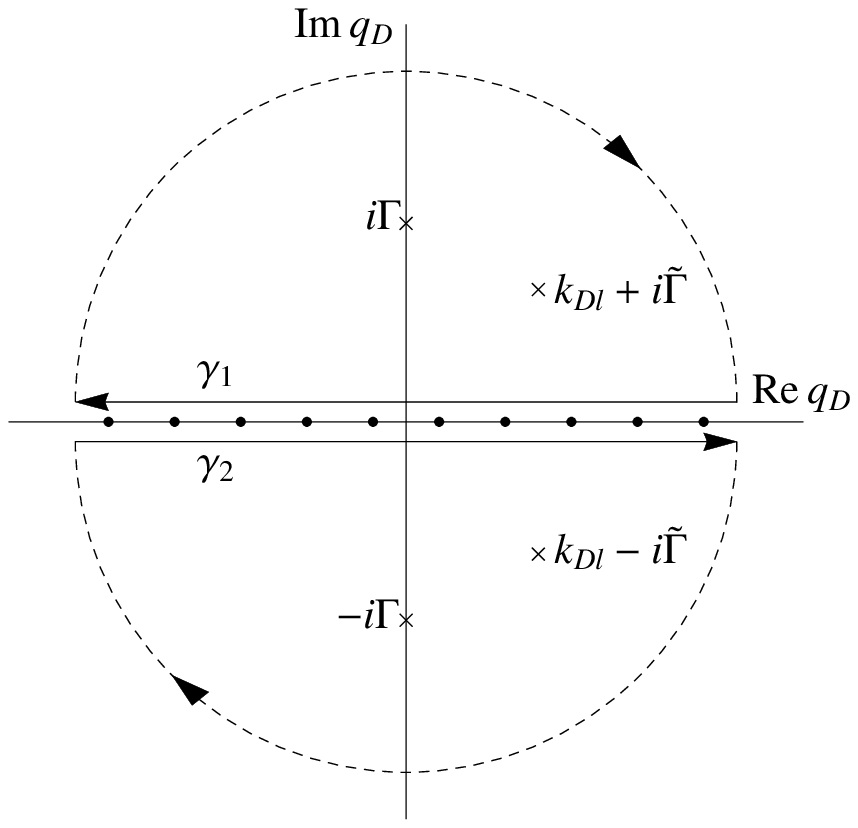}
\vspace*{-4cm}
\caption{\label{fg2}
The complex $q_D$-plane containing the integration paths $\gamma_1$ and $\gamma_2$.
The dots indicate the poles at the fermionic Matsubara frequencies.
The  four additional poles are shown as
 crosses  (see the text for further discussion).}
\end{figure}

Using the Cauchy residual theorem, the sum (\ref{eq18}) can be represented in the form
\eq{eq20}{ k_BT\sum_{n=-\infty}^\infty f(iq_{Dn}) &=
    -\!\!\!\int\limits_{\gamma_1\bigcup\gamma_2}\!\!
\frac{dq_D}{2\pi}\frac{f(iq_D)}{e^{i\frac{q_D}{k_BT}}+1}
}
where the integration contour in the complex $q_D$-plane shown in Fig.~\ref{fg2}
consists of the paths $\gamma_1$ and $\gamma_2$. The validity of Eq.~(\ref{eq20})
becomes evident when taking into account that the poles of the expression under
the integral are at the points $q_{Dn}=2\pi k_BT (n+1/2)$ shown as dots
in Fig.~\ref{fg2} and calculating the sum of the residues at these poles.

Substituting Eq.~(\ref{eq20}) in Eq.~(\ref{eq5}) and interchanging the order of
integrations, one obtains
\begin{eqnarray}
&&
\Pi^{\mu\nu}(ik_{Dl},\vk,T)=\frac{32\pi\al}{\v^2}\int\frac{d^{D-1}\q}{(2\pi)^{D-1}}
\label{eq21}\\
&&\times
\left[\int\limits_{\gamma_1}\frac{dq_D}{2\pi}\,\frac{1}{e^{i\frac{q_D}{k_BT}}+1}\,
\frac{Z^{\mu\nu}}{R}
+\int\limits_{\gamma_2}\frac{dq_D}{2\pi}\,\frac{1}{e^{i\frac{q_D}{k_BT}}+1}\,
\frac{Z^{\mu\nu}}{R}\right].
\nonumber
\end{eqnarray}
\noindent
Here, the integrand in the second term is decreasing in the lower half-plane.
To get the integrand in the first term decreasing in the upper half-plane,
in the integral along $\gamma_1$, we use the identity
\eq{eq22}{\frac{1}{e^{i\frac{q_D}{k_BT}}+1}=1-
\frac{1}{e^{-i\frac{q_D}{k_BT}}+1}.
}

Substituting it to Eq.~(\ref{eq21}), we bring the polarization tensor to the form
\begin{equation}
\Pi^{\mu\nu}(ik_{Dl},\vk,T)=\Pi_0^{\mu\nu}(ik_{Dl},\vk)+
\Delta_T\Pi^{\mu\nu}(ik_{Dl},\vk,T),
\label{eq23}
\end{equation}
\noindent
where
\begin{equation}
\Pi_0^{\mu\nu}(ik_{Dl},\vk)=-\frac{32\pi\al}{\v^2}\int_{-\infty}^{\infty}
\frac{dq_D}{2\pi}
\int\frac{d^{D-1}\q}{(2\pi)^{D-1}}\,\frac{Z^{\mu\nu}}{R}
\label{eq24}
\end{equation}
\noindent
and
\begin{eqnarray}
&&
\Delta_T\Pi^{\mu\nu}(ik_{Dl},\vk,T)=-\frac{32\pi\al}{\v^2}\int\frac{d^{D-1}\q}{(2\pi)^{D-1}}
\label{eq25}\\
&&\times
\left[\int\limits_{\gamma_1}\frac{dq_D}{2\pi}\,\frac{1}{e^{-i\frac{q_D}{k_BT}}+1}\,
\frac{Z^{\mu\nu}}{R}
-\int\limits_{\gamma_2}\frac{dq_D}{2\pi}\,\frac{1}{e^{i\frac{q_D}{k_BT}}+1}\,
\frac{Z^{\mu\nu}}{R}\right].
\nonumber
\end{eqnarray}
\noindent
Note that the sign minus in front of (\ref{eq24}) appeared because the direction of
the path $\gamma_1$ is against the real axis in the complex plane $q_D$.

The first term on the right-hand side of Eq.~(\ref{eq23}) given by Eq.~(\ref{eq24}) has the meaning of
the polarization tensor at zero temperature (till the moment it is calculated
at the bosonic Matsubara frequencies). As to the second term given by Eq.~(\ref{eq25}),
it explicitly depends on $T$ and has the meaning of the thermal correction.

We begin from calculation of the thermal correction. This can be done by closing the
integration paths $\gamma_1$ and $\gamma_2$ with the help of semicircles of the
infinitely large radii in the upper and lower half planes, respectively, and applying
again the Cauchy residue theorem. In the upper half-plane, there are two poles of the
function $Z^{\mu\nu}/R$ at the roots of $R$. These are
$q_D=i\Gamma(\q)$ and $q_D=i\tilde{\Gamma}(\q)+k_{Dl}$ where
$\Gamma$ and $\tilde{\Gamma}$ are defined in Eq.~(\ref{eq10}).
In the lower half-plane, the poles of the function $Z^{\mu\nu}/R$ are at
$q_D=-i\Gamma(\q)$ and $q_D=-i\tilde{\Gamma}(\q)+k_{Dl}$.
All these poles are shown in Fig.~\ref{fg2} as crosses.

Calculating the residues at all the four poles and taking into account that the
integrals along both semicircles vanish, we rewrite the thermal correction
(\ref{eq25}) as
\begin{widetext}
\begin{eqnarray}
&&
\Delta_T\Pi^{\mu\nu}(ik_{Dl},\vk,T)=\frac{16\pi\al}{\v^2}\int\frac{d^{D-1}\q}{(2\pi)^{D-1}}
\nonumber\\
&&~~\times
\sum_{\lambda=\pm 1}\left\{
\frac{Z^{\mu\nu}(q_D=i\lambda\Gamma)}{\Gamma\left(e^{\frac{\Gamma}{k_BT}}+1\right)
\left[(i\lambda\Gamma-k_{Dl})^2+\tilde{\Gamma}^2\right]}
\right.
\left.+
\frac{Z^{\mu\nu}(q_D=i\lambda\tilde{\Gamma}+k_{Dl})}{\tilde{\Gamma}
\left(e^{\frac{\tilde{\Gamma}}{k_BT}}+1\right)
\left[(i\lambda\tilde{\Gamma}+k_{Dl})^2+{\Gamma}^2\right]}
\right\}.
\label{eq26}
\end{eqnarray}
\noindent
When obtaining this equation, it was used that $\exp[-i\lambda k_{Dl}/(k_BT)]=1$
due to Eq.~(\ref{eq4}).

Equation (\ref{eq26}) can be further simplified because the integrand is symmetric
under the substitution $\q\rightarrow\tvk-\q$. Making this substitution and the
replacement $\lambda\rightarrow -\lambda$ in the second term of this equation,
one obtains

\begin{equation}
\Delta_T\Pi^{\mu\nu}(ik_{Dl},\vk,T)=\frac{16\pi\al}{\v^2}\int\frac{d^{D-1}\q}{(2\pi)^{D-1}}
\,\frac{1}{\Gamma\left(e^{\frac{\Gamma}{k_BT}}+1\right)}
\sum_{\lambda=\pm 1}
\frac{Z^{\mu\nu}(q_D=i\lambda\Gamma,\q)+
Z^{\mu\nu}(q_D=k_{Dl}-i\lambda\Gamma,\tvk-\q)}{(k_{Dl}-i\lambda\Gamma)^2
+\tilde{\Gamma}^2}.
\label{eq27}
\end{equation}
\end{widetext}

Taking into account that $\Gamma\sim |\q|$ when $|\q|\rightarrow\infty$,
it is seen that the integral in Eq.~(\ref{eq27}) converges exponentially fast for
any $D$.
Note that Eq.~(\ref{eq27}) is easily generalized for the case of graphene with
a nonzero chemical potential $\mu$. This is done by the replacement \cite{80a}
\begin{equation}
\frac{1}{e^{\frac{\Gamma}{k_BT}}+1}\longrightarrow\frac{1}{2}\left(
\frac{1}{e^{\frac{\Gamma+\mu}{k_BT}}+1}+
\frac{1}{e^{\frac{\Gamma-\mu}{k_BT}}+1}\right).
\label{eq27a}
\end{equation}
\noindent
Thus, the problem of convergence of the polarization tensor reduces to the
question of whether its zero-temperature part (\ref{eq24}) converges.

Note that the thermal correction in the form of Eq.~(\ref{eq27}) admits an
immediate analytic continuation to the real frequency axis by putting
$ik_{Dl}=k_0=\omega$ (compare with similar results obtained for the temperature
Green functions in Refs.~\cite{78,79} and with Ref.~\cite{8}).
In a similar way, the polarization tensor at zero temperature along the real
frequency axis is obtained from Eq.~(\ref{eq24}) by putting  $ik_{Dl}=k_0=\omega$
and $q_D=-iq_0$. With this substitution, it takes the form
\eq{eq28}{ \Pi_0^{\mu\nu}(k) &= i\frac{32\pi\alpha}{\v^2} \int  \frac{d^Dq}{(2\pi)^D}
        \frac{Z^{\mu\nu}(k,q)}{R(k,q)},
}
where $k=(k_0,\vk)$, $q=(q_0,\q)$ and  $d^Dq=dq_0\,d\q$.

According to Eqs.~(\ref{eq6}) and (\ref{eq7}), $Z^{\mu\nu}\sim q^2$ and $R\sim q^4$
in the limit $q^2\rightarrow\infty$. These simple power-counting
arguments show that the integral (\ref{eq28}) may
contain the ultraviolet divergences of the order of $q^{D-2}$, i.e., diverge linearly
and quadratically in three- and four-dimensional space-time, respectively.
Below we show how these expectations are modified by the gauge invariance of the
polarization tensor.

\section{Zero-temperature part and its analytic expression in {\boldmath$D$}
dimensions}

In this section, we calculate the zero-temperature polarization tensor (\ref{eq28})
in the case of $D$-dimensional space-time. For this purpose, we use the following
 representation for the propagators entering Eq.~(\ref{eq28}) \cite{80}:
\begin{eqnarray}
&&
\frac{1}{q^2-m^2+i0}=\frac{1}{i}\int_0^{\infty}dse^{is(q^2-m^2+i0)},
\label{eq29} \\
&&
\frac{1}{(q-\tk)^2-m^2+i0}=\frac{1}{i}\int_0^{\infty}dte^{it[(q-\tk)^2-m^2+i0]}.
\nonumber
\end{eqnarray}

For the momenta $q^{\nu^{\prime}}$ entering $\tzmn$ in Eq.~(\ref{eq7}), we use
\begin{equation}
q^{\,\nu^{\prime}}=\left.\frac{1}{i}\,\frac{\partial}{\partial\xi_{\nu^{\prime}}}
e^{iq^{\ga}\xi_{\ga}}\right|_{\xi=0}.
\label{eq30}
\end{equation}
\noindent
This substitution is made for all $q$ entering the function $\tzmn$, i.e.,
\begin{equation}
\tzmn(k,q)=\left.\tzmn\left(k,\frac{1}{i}\,\frac{\partial}{\partial\xi_{\nu^{\prime}}}
\right)e^{iq^{\ga}\xi_{\ga}}\right|_{\xi=0}.
\label{eq31}
\end{equation}

Substituting Eqs.~(\ref{eq6}) and (\ref{eq29}) in Eq.~(\ref{eq28}) with account of
the  definition of $R$ in Eq.~(\ref{eq7}) and using Eq.~(\ref{eq31}), the polarization
tensor at zero temperature is presented as
\begin{eqnarray}
&&
\Pi_{0}^{\mu\nu}(k)=\frac{32\pi\al}{i\v^2}\eta_{\mu^{\prime}}^{\,\mu}
\eta_{\nu^{\prime}}^{\,\nu}\int\frac{{d^Dq}}{(2\pi)^D}
\int_0^{\infty}ds
\label{eq32} \\
&&~~~~
\times
\int_0^{\infty}dt
\left.\tzmn\left(k,\frac{1}{i}\,\frac{\partial}{\partial\xi_{\nu^{\prime}}}
\right)e^{iM}\right|_{\xi=0},
\nonumber
\end{eqnarray}
where  the quantity $M$ is defined as
\begin{eqnarray}
&&
M = s(q^2-m^2+i0)
\nonumber \\
&&~~~~~~~
+t[(q-\tk)^2-m^2+i0]+i q\xi.
\label{eq33}
\end{eqnarray}

This expression for $M$ can be identically rewritten in the form
\begin{equation}
M = (s+t)\left[q+\frac{\xi-2t\tk}{2(s+t)^2}\right]^2
-\frac{\xi^2}{4(s+t)}+\frac{t}{s+t}\tk\xi+H,
\label{eq34}
\end{equation}
\noindent
where
\begin{equation}
H=\frac{st}{s+t}\tk^2-(s+t)m^2+i0.
\label{eq35}
\end{equation}

It is seen that only the first term in the expression (\ref{eq34}) for $M$ depends
on $q$. Then, the integration with respect to $q$ in Eq.~(\ref{eq32}) can be easily
performed. For this purpose, we use the well known formulas \cite{80}
{
\begin{eqnarray}
&&
\int_{-\infty}^{\infty}\frac{dq_0}{2\pi}\,e^{i(s+t)q_0^2}=
\frac{e^{i\frac{\pi}{4}}}{\sqrt{4\pi (s+t)}},
\label{eq36} \\
&&
\int_{-\infty}^{\infty}\frac{dq_j}{2\pi}\,
e^{-i(s+t)q_j^2}=
\frac{e^{-i\frac{\pi}{4}}}{\sqrt{4\pi (s+t)}},
\nonumber
\end{eqnarray}
\noindent
where $s+t>0$ and $j=1,\,2,\,\ldots,\,D-1$.}

Combining the necessary number of expressions in Eq.~(\ref{eq36}), for the
$D$-dimensional space-time one obtains
{
\begin{equation}
\int\frac{d^Dq}{(2\pi)^D}\,e^{i(s+t)q^2}=
\frac{e^{i\frac{\pi}{4}(2-D)}}{[4\pi (s+t)]^{\frac{D}{2}}}.
\label{eq37}
\end{equation}}

By applying Eq.~(\ref{eq37}) with a necessary shift of the integration variable
$q$  in Eqs.~(\ref{eq32}) and (\ref{eq34}), one obtains
\begin{eqnarray}
&&
\Pi_0^{\mu\nu}(k)=\frac{32\pi\al}{i\v^2}\,e^{i\frac{\pi}{4}(2-D)}
\eta_{\mu^{\prime}}^{\,\mu}
\eta_{\nu^{\prime}}^{\,\nu}\int_0^{\infty}\!\!\!ds\int_0^{\infty}\!
\frac{dt}{[4\pi(s+t)]^{D/2}}
\nonumber \\
&&~~~
\times \ozmn\,e^{iH},
\label{eq38}
\end{eqnarray}
\noindent
where
\begin{equation}
\ozmn=\tzmn\left.\left(\tk,\frac{1}{i}\frac{\pa}{\pa\xi_{\nu^{\prime}}}\right)
\exp\left[\frac{i}{4(s+t)}(4t\tk\xi-\xi^2)\right]\right|_{\xi=0}.
\label{eq39}
\end{equation}

The functional form of the quantity $\tzmn$ is presented in the first line of
Eq.~(\ref{eq7}). It is seen that, in order to calculate the quantity (\ref{eq39}),
one should find how the operators obtained from $q^{\mu^{\prime}}$,
$q^{\mu^{\prime}}q^{\nu^{\prime}}$, and $q^2$ by the replacement of $q^{\mu^{\prime}}$
with $-i\pa/\pa\xi_{\mu^{\prime}}$ act on the exponent in Eq.~(\ref{eq39}).
As an example,
\begin{eqnarray}
&&
\frac{1}{i}\frac{\pa}{\pa\xi_{\mu^{\prime}}}
\exp\left[\frac{i}{4(s+t)}(4t\tk\xi-\xi^2)\right]=
\frac{2t\tk^{\mu^{\prime}}-\xi^{\mu^{\prime}}}{2(s+t)}
\nonumber \\
&&~~~~~\times
\exp\left[\frac{i}{4(s+t)}(4t\tk\xi-\xi^2)\right].
\label{eq40}
\end{eqnarray}

By putting here $\xi=0$, one finds
\begin{equation}
\left.\frac{1}{i}\frac{\pa}{\pa\xi_{\mu^{\prime}}}
\exp\left[\frac{i}{4(s+t)}(4t\tk\xi-\xi^2)\right]\right|_{\xi=0}=
\frac{t}{s+t}\tk^{\mu^{\prime}}.
\label{eq41}
\end{equation}

In a similar way, calculating the remaining derivatives and putting $\xi=0$ in the
obtained results, we arrive at
\begin{eqnarray}
&&
\left.\frac{1}{i}\frac{\pa}{\pa\xi_{\mu^{\prime}}}
\frac{1}{i}\frac{\pa}{\pa\xi_{\nu^{\prime}}}
\exp\left[\frac{i}{4(s+t)}(4t\tk\xi-\xi^2)\right]\right|_{\xi=0}
\nonumber \\
&&~~~~~~~
=
-\frac{\gmn}{2i(s+t)}+\frac{t^2}{(s+t)^2}\tk^{\mu^{\prime}}
\tk^{\nu^{\prime}},
\nonumber \\
&&
\left.\frac{1}{i}\frac{\pa}{\pa\xi^{\mu^{\prime}}}
\frac{1}{i}\frac{\pa}{\pa\xi_{\mu^{\prime}}}
\exp\left[\frac{i}{4(s+t)}(4t\tk\xi-\xi^2)\right]\right|_{\xi=0}
\nonumber \\
&&~~~~~~~
=
-\frac{D}{2i(s+t)}+\frac{t^2}{(s+t)^2}\tk^2,
\label{eq42}
\end{eqnarray}
where we accounted for $g_{\mu\nu}g^{\mu\nu}=D$.

Using Eqs.~(\ref{eq7}), (\ref{eq41}), and (\ref{eq42}), we bring Eq.~(\ref{eq39})
to the form
\begin{eqnarray}
&&
\ozmn=\gmn\frac{D-2}{2i(s+t)}-\frac{ts}{(s+t)^2}(2\tk^{\mu^{\prime}}
\tk^{\nu^{\prime}}-\gmn\tk^2)
\nonumber\\
&&~~~~~~~~~~~~~
+\gmn m^2.
\label{eq43}
\end{eqnarray}

It is convenient to rewrite the polarization tensor (\ref{eq38}) in terms of new
integration variables $\rho$ and $\lambda$ defined as
\begin{equation}
s=\rho\lambda,\qquad t=(1-\rho)\lambda
\label{eq44}
\end{equation}
\noindent
so that
\begin{equation}
\rho=\frac{s}{s+t},\qquad \lambda=s+t,
\label{eq45}
\end{equation}
\noindent
 where $\rho$ is the so-called Feynman parameter (frequently denoted by $x$).

 It is easily seen that {
\begin{equation}
\int_{0}^{\infty}ds\int_{0}^{\infty}dt\,g(s,t)=
\int_{0}^{1}d\rho\int_{0}^{\infty}\lambda d\lambda\,
g\Big(\rho\lambda,(1-\rho)\lambda\Big),
\label{eq46}
\end{equation}
\noindent
where the factor $\lambda$ in Eq.~(\ref{eq46}) comes from the Jacobian.}

In terms of the variables (\ref{eq44}), the quantities $\ozmn$ from Eq.~(\ref{eq43})
and $H$  from Eq.~(\ref{eq35}) take the form
\begin{eqnarray}
&&
\ozmn=\gmn\frac{D-2}{2i\lambda}-2\rho(1-\rho)\tk^{\mu^{\prime}}
\tk^{\nu^{\prime}}
\nonumber\\
&&~~~~~~~~~~
+\gmn[\rho(1-\rho)\tk^2+m^2],
\label{eq47}\\
&&
H=\lambda[\rho(1-\rho)\tk^2-m^2+i0]\equiv\lambda H_1(\rho).
\nonumber
\end{eqnarray}

Then, the polarization tensor (\ref{eq38}) is given by
\begin{eqnarray}
&&
\Pi_0^{\mu\nu}(k)=\frac{32\pi\al}{i\v^2}\,e^{i\frac{\pi}{4}(2-D)}
\eta_{\mu^{\prime}}^{\,\mu}
\eta_{\nu^{\prime}}^{\,\nu}\int_0^{1}\!\!\!d\rho\int_0^{\infty}\!
\frac{d\lambda}{(4\pi)^{D/2}}
\nonumber \\
&&~~~
\times \lambda^{1-\frac{D}{2}}\ozmn\,e^{i\lambda H_1(\rho)},
\label{eq48}
\end{eqnarray}
\noindent
where $\ozmn$ and $H_1$ are defined in Eq.~(\ref{eq47}). Note that the limit
of large momenta corresponds to small $\lambda$.

The integral over $\lambda$ in Eq.~(\ref{eq48}) can be calculated using the
formula \cite{81}
\begin{equation}
\int_0^{\infty}d\lambda e^{i\lambda H_1(\rho)}\lambda^{w-1}=[-iH_1(\rho)]^{-w}
\Gamma(w),
\label{eq49}
\end{equation}
\noindent
where $\Gamma(w)$ is the gamma function. Note that the integral on the left-hand
side of Eq.~(\ref{eq49}) is equal to the gamma function only under the conditions
${\rm Re}(-iH_1)>0$ and ${\rm Re}\,w>0$. The first of them is satisfied due to the
presence of $i0$ in Eq.~(\ref{eq47}). Below we apply Eq.~(\ref{eq49}) for the
space-time with ${\rm Re}D<2$ where ${\rm Re}\,w>0$. The results for the cases
${\rm Re}D>2$ are
obtained by the standard analytic continuation (see the next section for the
differences between the cases $D=3$ or $D=4$).

Using Eq.~(\ref{eq49}) in Eq.~(\ref{eq48}), one finds
\begin{widetext}
\begin{eqnarray}
&&
\int_0^{\infty}\!
\frac{d\lambda}{(4\pi)^{D/2}}
\lambda^{1-\frac{D}{2}}\ozmn\,e^{i\lambda H_1(\rho)}=
\gmn\frac{D-2}{2i(4\pi)^{D/2}}\Gamma\left(1-\frac{D}{2}\right)
[-iH_1(\rho)]^{\frac{D}{2}-1}
\label{eq50}\\
&&~~~
+\frac{1}{(4\pi)^{D/2}}\left\{-2\rho(1-\rho)\tk^{\mu^{\prime}}\tk^{\nu^{\prime}}
 +\gmn[\rho(1-\rho)\tk^2+m^2]\right\}
\Gamma\left(2-\frac{D}{2}\right)
[-iH_1(\rho)]^{\frac{D}{2}-2}.
\nonumber
\end{eqnarray}
\end{widetext}

Using the property
\begin{equation}
\Gamma(z)=\frac{\Gamma(z+1)}{z},
\label{eq51}
\end{equation}
\noindent
the integral (\ref{eq50}) can be rewritten in a simpler form
\begin{eqnarray}
&&
\int_0^{\infty}\!
\frac{d\lambda}{(4\pi)^{D/2}}
\lambda^{1-\frac{D}{2}}\ozmn\,e^{i\lambda H_1(\rho)}=
\frac{2}{(4\pi)^{D/2}}\rho(1-\rho)
\nonumber \\
&&~
\times (\gmn\tk^2-\tk^{\mu^{\prime}}
\tk^{\nu^{\prime}})
\Gamma\left(2-\frac{D}{2}\right)
[-iH_1(\rho)]^{\frac{D}{2}-2}.
\label{eq52}
\end{eqnarray}

Inserting Eq.~(\ref{eq52}) in Eq.~(\ref{eq48}), we arrive at
\begin{eqnarray}
&&
\Pi_0^{\mu\nu}(k)=\frac{64\pi\al}{i\v^2}\,e^{i\frac{\pi}{4}(2-D)}
\frac{\eta_{\mu^{\prime}}^{\,\mu}\eta_{\nu^{\prime}}^{\,\nu}}{(4\pi)^{D/2}}
\Gamma\left(2-\frac{D}{2}\right)
\label{eq53} \\
&&~
\times
(\gmn\tk^2-\tk^{\mu^{\prime}}\tk^{\nu^{\prime}}) \int_0^1d\rho\rho(1-\rho)
[-iH_1(\rho)]^{\frac{D}{2}-2}.
\nonumber
\end{eqnarray}
\noindent
{}From the tensor structure of Eq.~(\ref{eq53}) it becomes evident that for
$\Pi_0^{\mu\nu}$ the transversality condition (\ref{eq14}) is satisfied as it
must be for both the zero-temperature part of the polarization tensor and for the
thermal correction to it.

The analytic continuations of Eq.~(\ref{eq53}) to the cases of $D=4$ and $D=3$ are considered in the next section.

\section{Three- and four-dimensional space-times}

We begin with the case of four-dimensional space-time $D=4$.
{Keeping in mind the necessity of regularization, let us put
$D=4-2\ve$, where $\ve$ vanishes when $D$ goes to 4.} In this case
Eq.~(\ref{eq53}) takes the form
\begin{eqnarray}
&&
\Pi_{0,\ve}^{\mu\nu}(k)=-\frac{4\al}{\pi\v^2}
\eta_{\mu^{\prime}}^{\,\mu}\eta_{\nu^{\prime}}^{\,\nu}
\Gamma\left(\ve\right)
(\gmn\tk^2-\tk^{\mu^{\prime}}
\tk^{\nu^{\prime}})
\nonumber \\
&&~~~
\times  \int_0^1d\rho\rho(1-\rho)
H_1^{-\ve}(\rho).
\label{eq54}
\end{eqnarray}

In fact the gamma function on the right-hand side of Eq.~(\ref{eq54}) can be
analytically continued to the entire plane of complex $\ve$ with exception of the
poles at $\ve=0,\,-1,\,-2,\,\ldots\,$. This allows to perform the dimensional
regularization of the polarization tensor (\ref{eq54})
{and subsequent renormalization}
by subtracting the pole
contribution in the form of $1/\ve$.

To do so, we expand the gamma function according to \cite{81}
\begin{equation}
\Gamma(\ve)=\frac{1}{\ve}-\gamma+O(\ve),
\label{eq55}
\end{equation}
\noindent
where $\gamma$ is the Euler constant. The factor $H_1^{-\ve}$   is represented as
\begin{equation}
H_1^{-\ve}=\exp\left[\ln\left(\frac{H_1(\rho)}{C}\right)^{-\ve}\right]
=1-\ve\ln\frac{H_1(\rho)}{C}+O(\ve^2),
\label{eq56}
\end{equation}
\noindent
where $C$ is an arbitrary constant with the dimension of $H_1$.

Substituting Eqs.~(\ref{eq55}) and (\ref{eq56}) in Eq.~(\ref{eq54}), one obtains
\begin{eqnarray}
&&
\Pi_{0,\ve}^{\mu\nu}(k)=-\frac{4\al}{\pi\v^2}
\eta_{\mu^{\prime}}^{\,\mu}\eta_{\nu^{\prime}}^{\,\nu}
(\gmn\tk^2-\tk^{\mu^{\prime}}
\tk^{\nu^{\prime}})
\nonumber \\
&&~~~
\times  \int_0^1\!\!d\rho\rho(1-\rho)
\left(\frac{1}{\ve}-\ln\frac{H_1(\rho)}{{C^{\prime}}}\right),
\label{eq57}
\end{eqnarray}
{where $C^{\prime}=Ce^{-\gamma}$.}

It is convenient to rewrite this result in the form
\begin{equation}
\Pi_{0,D\to4}^{\mu\nu}(k)=
\eta_{\mu^{\prime}}^{\,\mu}\eta_{\nu^{\prime}}^{\,\nu}
(\gmn\tk^2-\tk^{\mu^{\prime}}
\tk^{\nu^{\prime}})\Pi_4(k^2),
\label{eq58}
\end{equation}
\noindent
where
\begin{equation}
\Pi_4(k^2)=\frac{4\al}{\pi\v^2}
\int_0^1\!\!d\rho\rho(1-\rho)
\left(\frac{2}{D-4}+\ln\frac{H_1(\rho)}{{C^{\prime}}}\right),
\label{eq59}
\end{equation}
\noindent
and, in accordance with Eq.~(\ref{eq47}),
\begin{equation}
H_1(\rho)=\rho(1-\rho)\tk^2-m^2+i0.
\label{eq60}
\end{equation}

The renormalization in quantum electrodynamics with $D=4$ consists in discarding
the pole term in Eq.~(\ref{eq59}) which corresponds to the logarithmic ultraviolet
divergence. This divergence is by two powers less than it follows from a simple power
counting for $D=4$ discussed in the end of Sec.~II. The decrease in the divergence
power is the result
of transversality (gauge invariance) of the polarization tensor ensured by the
tensor structure of Eq.~(\ref{eq58}). By imposing the normalization condition
$\Pi_4^{\rm ren}(k^2=0)=0$ (which is justified by the general theory of renormalization
in QED), one can fix the arbitrary constant
{$C^{\prime}=-m^2$} and arrive at
\begin{equation}
\Pi_4^{\rm ren}(k^2)=\frac{4\al}{\pi\v^2}
\int_0^1\!\!d\rho\rho(1-\rho)
\ln\left[1-\rho(1-\rho)\frac{\tk^2}{m^2}\right].
\label{eq61}
\end{equation}
\noindent
This is the well known result of the standard QED \cite{80} if we put $\v=1$ and
consider one Dirac point in place of two as for graphene.

Now we pass to the case $D=3$, i.e., to the polarization tensor of graphene at
zero temperature. In this case Eq.~(\ref{eq53}) takes the form
\begin{eqnarray}
&&
\Pi_{0,D=3}^{\mu\nu}(k)=-\frac{8\al}{\sqrt{\pi}\v^2}
\eta_{\mu^{\prime}}^{\,\mu}\eta_{\nu^{\prime}}^{\,\nu}
\Gamma\left(\frac{1}{2}\right)
(\gmn\tk^2-\tk^{\mu^{\prime}}
\tk^{\nu^{\prime}})
\nonumber \\
&&~~~
\times  \int_0^1d\rho\frac{\rho(1-\rho)}{\sqrt{-H_1(\rho)}}.
\label{eq62}
\end{eqnarray}

This equation, similar to Eq.~(\ref{eq54}), is obtained by the analytic continuation
of Eq.~(\ref{eq53}). However, as opposed to Eq.~(\ref{eq54}),
it is finite and does not contain the pole terms. Thus, no subtraction of infinities
is needed for obtaining the final physical result, i.e., the polarization tensor of
graphene behaves like that in the truly three-dimensional QED
which is the super-renormalizable theory (as mentioned especially in \cite{1}), unlike the
 standard theory in four dimensions which is ``only" renormalizable.

Using the same representation as in Eq.~(\ref{eq58})
\begin{equation}
\Pi_{0,D=3}^{\mu\nu}(k)=
\eta_{\mu^{\prime}}^{\,\mu}\eta_{\nu^{\prime}}^{\,\nu}
(\gmn\tk^2-\tk^{\mu^{\prime}}
\tk^{\nu^{\prime}})\Pi_3(k^2),
\label{eq63}
\end{equation}
\noindent
one obtains from Eq.~(\ref{eq62})
\begin{equation}
\Pi_3(k^2)=-\frac{8\al}{\v^2}\int_0^1\!\!d\rho
\frac{\rho(1-\rho)}{\sqrt{m^2-\rho(1-\rho)\tk^2}}.
\label{eq64}
\end{equation}

The last integral is easily calculated \cite{81}. Thus,
\begin{eqnarray}
&&
\Pi_3(k^2)=-\frac{4\al}{\v^2\tk^2}\left[
\vphantom{\frac{1}{\sqrt{m^2-\rho(1-\rho)\tk^2}}}
-m+\frac{4m^2+\tk^2}{4}\right.
\nonumber\\
&&~~~~~~~~
\times\left.
\int_0^1\!\!d\rho
\frac{1}{\sqrt{m^2-\rho(1-\rho)\tk^2}}\right].
\label{eq65}
\end{eqnarray}
\noindent
Using the most convenient expressions for this integral in different regions of
parameters, for $\tk^2<0$ we obtain
\begin{equation}
\Pi_3(k^2)=\frac{2\al}{\v^2\tk^2}\left(2m-\frac{4m^2+\tk^2}{\sqrt{-\tk^2}}
{\rm arctan}\frac{\sqrt{-\tk^2}}{2m}\right).
\label{eq66}
\end{equation}

Under the conditions $\tk^2>0$, $2m>\sqrt{\tk^2}$ we have
\begin{equation}
\Pi_3(k^2)=\frac{2\al}{\v^2\tk^2}\left(2m-\frac{4m^2+\tk^2}{\sqrt{\tk^2}}
{\rm arctanh}\frac{\sqrt{\tk^2}}{2m}\right).
\label{eq67}
\end{equation}

Finally, under the conditions $\tk^2>0$, $2m<\sqrt{\tk^2}$ one obtains
\begin{eqnarray}
&&
\Pi_3(k^2)=\frac{2\al}{\v^2\tk^2}\left[2m-\frac{4m^2+\tk^2}{\sqrt{\tk^2}}
\right.
\nonumber \\
&&~~~~~
\times \left.\left(
{\rm arctanh}\frac{2m}{\sqrt{\tk^2}}+i\frac{\pi}{2}\right)\right].
\label{eq68}
\end{eqnarray}
\noindent
Note that there is a threshold at $\sqrt{\tk^2}=2m$.

The two convenient independent quantities characterizing the polarization tensor
are $\Pi^{00}$ and ${\rm tr}\Pi^{\mu\nu}=g_{\mu\nu}\Pi^{\mu\nu}$.
Using Eq.~(\ref{eq63}), these are given by
\begin{eqnarray}
&&
\Pi_{0,D=3}^{00}(k)=-\v^2\vk^2\Pi_3(k^2),
\nonumber \\
&&
{\rm tr} \Pi_{0,D=3}^{\mu\nu}(k)= \v^2(k^2+\tk^2)\Pi_3(k^2),
\label{eq69}
\end{eqnarray}
\noindent
where $\Pi_3(k^2)$ is defined in Eqs.~(\ref{eq66})--(\ref{eq68}) for different
regions of the involved parameters. From Eq.~(\ref{eq63}) it is seen that if the
mass shell equation $k_0^2-\vk^2=0$ is satisfied, it holds $\Pi^{\mu\nu}(k_0=0)=0$.

{
Equations (\ref{eq63}) and (\ref{eq66})--(\ref{eq69}) coincide with the results of
Refs.~\cite{15,30,36} for the polarization tensor of graphene at zero temperature.
It should be added also that
the equivalent results \cite{27} were found in the literature by the method of
correlation functions in the random-phase approximation \cite{81a,81b,81c,81d}.
 The obtained results are unique and neither $\Pi_{0}^{00}$ nor
${\rm tr}\Pi_{0}^{\mu\nu}$ can be modified in any way.
As to the thermal correction to the polarization tensor $\Delta_{T}\Pi^{\mu\nu}$, in Sec.~II it was shown that
it is finite for any $D$ and defined uniquely. Because of this, it is not the
subject of regularization which refers to only the zero-temperature case.
}

We underline that Eqs.~(\ref{eq63}) and (\ref{eq66})--(\ref{eq68}) for the
polarization tensor of graphene at zero temperature, where the Fermi velocity $\v$
is put equal to unity, are in agreement with the well known results of Refs.~\cite{0,1}
obtained long ago in the framework of the standard (2+1)-dimensional QED (the extra
factor 2 is explained by the presence of two Dirac points for graphene).

\section{Conclusions and discussion}

In the foregoing, we have analyzed the problem of convergence of the polarization
tensor of graphene in the framework of the Dirac model. This is an interesting example
regarding application of methods of the low-dimensional thermal QFT
to a material of big practical importance. Although in the framework of QFT
the polarization tensor of graphene is described by a simple one-loop diagram,
which was calculated long ago, there are contradictory statements in the literature
mentioned in Sec.~I concerning its convergence, the necessity of its regularization
and validity of the obtained results. Taking into account that the quantum field
theoretical approach to the polarization tensor of graphene suggests the most direct
and fundamental way for investigating the electrical conductivity and reflectance
of graphene, as well as the Casimir effect in graphene systems, it seems necessary
to clarify all the raised points.

For this purpose, we have performed a detailed calculation of the polarization tensor
of graphene and analyzed its analytic properties as a function of the number of
space-time dimensions. It is underlined that this tensor consists of the zero-temperature
part plus the thermal correction. In so doing, the thermal correction is represented
as an integral which converges in the space-time of any dimensionality. Thus, the
question of regularization is irrelevant to the thermal correction and may be raised
only with respect to the zero-temperature part of the polarization tensor.

For experts in QFT, calculation of the polarization tensor in the
framework of (2+1)-dimensional QED is a rather simple exercise.
Because of this, in the classical papers \cite{0,1} the results of this calculation
were presented without derivation. In Refs.~\cite{15,30,36}, again with no detailed
derivation, these results were modified for the case of graphene by taking into
account the presence of two fundamental velocities.

As discussed in Sec.~I, some of the theoretical predictions made using the quantum
field theoretical polarization tensor (and especially its trace) are in disagreement
with those found with the polarization tensor derived by the Kubo formula.
To bring both tensors in agreement, an alternative regularization procedure was
suggested \cite{75} by imposing an artificial additional condition irrelevant to
the rigorous formalism of quantum field theory.

Our detailed analysis of the convergence of the polarization tensor in
$D=(2+1)$-dimensional space-time shows that, although it is formally represented by
a divergent integral, its finite value is obtained by the analytic continuation.
 In so doing, one need not to discard any pole terms
which do not appear in the case $D=3$,
{ i.e., the renormalization is not needed.}
Just this was meant in Refs.~\cite{28,29,30}
stating that for $D=3$ the ultraviolet divergences do not appear. After putting the
Fermi velocity equal to the speed of light, our results for the zero-temperature
polarization tensor are found in agreement with the well known results of
Refs.~\cite{0,1}. If the two fundamental velocities are present, our results
coincide with those given for graphene in Refs.~\cite{15,30,36}.

We remind that the situation is different in the case of the standard QED with
$D=3+1$. In this case, the zero-temperature polarization tensor is also obtained by
the analytic continuation. However, for obtaining the
finite result, it is necessary to discard the pole term which arises for $D=4$.
This pole corresponds to the ultraviolet divergence deleted by means of the
renormalization procedure,
{ which must be performed after a regularization.}
Therefore, there is a principal difference between the
character of divergences of the polarization tensor for the three- and four-dimensional
space-times. In both cases, however, the final results, obtained by the analytic
continuation from the case of lower dimensionality and (for $D=4$ only) by discarding
the pole term and using the normalization condition,
are unique and not a subject to any modification.

It is also necessary to stress also that the presence of a double pole at zero frequency
in the transverse dielectric permittivity of graphene proven by using the polarization
tensor \cite{42} plays a decisive role in reaching an agreement between theory and
measurements of the Casimir force in graphene systems \cite{51,52,53,54}. It is well
known that for metallic test bodies the theoretical predictions are in agreement with the
results of numerous precise experiments on measuring the Casimir force only if the
response of metals to the low-frequency electromagnetic field is described by the
dissipationless plasma model possessing a double pole at zero frequency \cite{82,83}.
This problem was considered as a failure of the dissipative Drude model, possessing
the single pole at zero frequency, in the region of transverse electric evanescent
waves \cite{84}. Thus, a prediction of the double pole in the transverse dielectric
permittivity of graphene in the framework of quantum field theory, as opposed to the
Kubo formula, is in favor of the former.

To conclude, the analysis performed in this paper opens opportunities for a wider
use of quantum field theoretical methods for investigation of the properties of
graphene and other novel materials.

\section*{Acknowledgments}
M.B.\ thanks I.\ G.\ Pirozhenko for valuable discussions.
G.L.K.\ and V.M.M.\ are grateful to M.\ I.\ Eides for useful comments.
The work by G.L.K.\ and V.M.M.~was partially funded by the
Ministry of Science and Higher Education of Russian Federation
(``The World-Class Research Center: Advanced Digital Technologies,"
contract No. 075-15-2022-311 dated April 20, 2022). The research
of V.M.M.\ was also partially carried out in accordance with the Strategic
Academic Leadership Program ``Priority 2030" of the Kazan Federal
University.


\end{document}